\documentclass[12pt]{iopart}

\usepackage{iopams}
\usepackage{graphicx}

\begin{document}

\title[The asymmetric quantum Rabi model and generalised P\"oschl-Teller potentials]{The asymmetric quantum Rabi model and generalised P\"oschl-Teller potentials}

\author{Kai-Long Guan$^{1}$, Zi-Min Li$^{2}$, Clare Dunning$^{3}$ and Murray T Batchelor$^{1,2,4}$}

\address{$^{1}$Centre for Modern Physics, Chongqing University, Chongqing 400044, China}

\address{$^{2}$Department of Theoretical Physics, 
Research School of Physics and Engineering, Australian National University, Canberra, ACT 2601, Australia}

\address{$^{3}$School of Mathematics, Statistics and Actuarial Science, University of Kent, Canterbury CT2 7NZ, UK}

\address{$^{4}$Mathematical Sciences Institute, Australian National University, Canberra ACT 2601, Australia}

\ead{batchelor@cqu.edu.cn}

\begin{abstract}
Starting with the Gaudin-like Bethe ansatz equations associated with the quasi-exactly solved (QES) exceptional points of the asymmetric quantum Rabi model (AQRM) 
a spectral equivalence is established with QES hyperbolic Schr\"odinger potentials on the line.
This leads to particular QES P\"oschl-Teller potentials. 
The complete spectral equivalence is then established between the AQRM and generalised P\"oschl-Teller potentials. 
This result extends a previous mapping between the symmetric quantum Rabi model and a QES P\"oschl-Teller potential.
The complete spectral equivalence between the two systems suggests that the physics of the generalised P\"oschl-Teller potentials may also be 
explored in experimental realisations of the quantum Rabi model. 
\end{abstract}

\section{Introduction}

The quantum Rabi model \cite{Rabi,JC} describes the fundamental interaction between a two-level atom and a single-mode bosonic field. 
There are a number of reasons for the recent growth of interest in the quantum Rabi model, from the perspectives of both mathematics and physics \cite{intro,review}.
Briefly stated, this is because experiments are now able to push into the coupling regimes beyond which the simpler Jaynes-Cummings model \cite{JC} no longer applies, 
with prospects for novel regimes of light-matter interactions. 
The analytic solution of the quantum Rabi model \cite{Braak} has also inspired further interest in the analysis of this class of models.

The asymmetric version of the quantum Rabi model of interest here is a particular generalisation of the Rabi model described by the hamiltonian 
\begin{equation}
H= \Delta \, \sigma_z + \epsilon\,\sigma_x + \omega \, a^{\dagger}a + g \, \sigma_x(a^{\dagger}+a)  . \label{ham}
\end{equation}
Here $\sigma_x$ and $\sigma_z$ are Pauli matrices for a two-level system with level splitting $\Delta$.
The single-mode bosonic field is described by the creation and destruction operators $a^\dagger$  and $a$ with 
$[a, a^\dagger] = 1$ and frequency $\omega$.
The interaction between the matter and light systems is via the coupling $g$. 
The additional term $\epsilon \, \sigma_x$ breaks the $Z_2$ symmetry (parity) of the Rabi model.  
It allows tunnelling between the two atomic states. 
The asymmetric version of the quantum Rabi model is relevant to the description of various hybrid mechanical systems \cite{Heun2,hybrid}. 
Moreover, the asymmetric quantum Rabi model (AQRM) is unitarily equivalent to the effective circuit QED hamiltonian describing a flux qubit \cite{cQED}
\begin{equation}
H_{\mathrm{cQED}} = \frac{1}{2} \Omega  \, \sigma_z + \omega \, a^{\dagger} a + g (\cos \theta \, \sigma_x - \sin \theta \, \sigma_y)  (a + a^{\dagger})   \, ,
\end{equation}
with $\Delta = \frac12 {\Omega} \sin \theta$ and  $\epsilon = \frac12 {\Omega} \cos \theta$.

The AQRM has been solved only relatively recently.
There have been two approaches: (i) 
by mapping the problem to the Bargmann space of analytic functions \cite{Braak}, and 
(ii) by using the Bogoliubov operator method \cite{Chen}. 
Using the former approach explicit expressions have been obtained \cite{Heun2,others2} for the wavefunction 
in terms of confluent Heun functions.
Of particular relevance here is the fact that the energy spectrum of the AQRM, 
although possessing no parity symmetry, still includes both regular and exceptional parts. 
The full eigenspectrum can be determined from the analytical solution. 
The exceptional parts, known as Juddian isolated exact solutions \cite{Judd}, 
can be systematically found from the conditions under which the confluent 
Heun functions are terminated as finite polynomials \cite{Heun2,LB1}. 
The eigenvalues are simply those of a shifted oscillator, 
however with the system parameters satisfying constraint polynomials which become  
increasingly complicated for higher energy levels.
A significantly deep understanding of the constraint polynomials has recently been obtained, 
paving the way for the general proof of crossing points in the energy spectrum when  $\epsilon/\omega \in \case12 \mathbb{Z}$
 \cite{Wakayama1,Wakayama2}. 
These crossing points become conical intersection points when the energy surface is considered in the $(g,\epsilon)$ parameter space \cite{BLZ}.

Our starting point is with algebraic Bethe ansatz equations characterising the 
exceptional part of the eigenspectrum of the AQRM.
These equations were obtained \cite{LB1} following the connection made \cite{Koc,Zhang} between the quantum Rabi model 
and the theory of quasi-exactly solved (QES) models.
For this reason the quantum Rabi model has been called a QES model, but given there is an analytic solution for the full eigenspectrum, 
such a label seems not entirely appropriate \cite{review,BZ}. 
Nevertheless, the identification of a QES sector of the Rabi model is important. 
The notion of QES comes from quantum mechanics where there exist potentials for which it is possible to find a finite number of exact 
eigenvalues and associated eigenfunctions in a relatively simple closed algebraic form \cite{QESa,QESb}.
In particular, there is a connection between QES Schr\"odinger potentials and Gaudin-like Bethe ansatz equations \cite{DHL}. 
Here we complete this circle by establishing the explicit connection between the algebraic QES part of the eigenspectrum of the AQRM and QES hyperbolic Schr\"odinger potentials. 
In doing so, we obtain a specific QES generalisation of the well known P\"oschl-Teller potential \cite{PT}.
The P\"oschl-Teller potential has appeared in many areas of physics, including quantum many-body systems, quantum wells, black holes   
and optical waveguides.\footnote{See, e.g., \cite{HP2017} and references therein.}

In units of $\hbar = 2m =1$, the relevant results \cite{DHL} satisfying the one-dimensional Schr\"odinger equation
\begin{equation}
- \frac{d^2 \Psi(x)}{dx^2} + V(x) \Psi(x) = {\cal E} \Psi(x) , 
\label{SE}
\end{equation}
are the wavefunction 
\begin{eqnarray}
&& \Psi(x) = (\cosh x \, - 1)^{-(B/2+1/4)} (\cosh x \, +1)^{-(C/2+1/4)} \nonumber \\
&& \qquad \qquad  \times \exp\left( \frac{A\gamma}{4} \cosh x \right) \prod_{j=1}^M \left( \frac{\gamma}{2} \cosh x \, + v_j \right), 
\label{psi}
\end{eqnarray}
with Schr\"odinger potential 
\begin{eqnarray}
&& V(x; A,B,C,\gamma) = M ( M -1 - B - C  + \frac{A\gamma}{2} \cosh x ) + \frac14 (B+C+1)^2  \nonumber \\
&& \qquad + \frac{A^2 \gamma^2}{16} \sinh^2 x + \frac{A \gamma }{4}{(C-B)}- \frac{A\gamma}{4}{(B+C)} \cosh x \nonumber \\
&& \qquad + \frac{(2B+1)(2B+3)}{8(\cosh x \, -1)} - \frac{(2C+1)(2C+3)}{8(\cosh x \, +1)} . 
\label{pot}
\end{eqnarray}
The general form of the algebraic Bethe ansatz equations is 
\begin{equation}
A + \frac{B}{v_j + \frac{1}{2} \gamma} + \frac{C}{v_j - \frac{1}{2}\gamma} = \sum_{k \ne j}^M \frac{2}{v_j - v_k} , 
\label{BAE}
\end{equation}
with 
\begin{equation}
{\cal E} = A \sum_{j=1}^M v_j .
\label{calE}
\end{equation}
The procedure for a specific hamiltonian is to identify the parameters $A, B, C$ and $\gamma$ from the corresponding set of Bethe ansatz equations (\ref{BAE}), 
from which the wavefunction and Schr\"odinger potential $V(x; A,B,C,\gamma)$ follow.
In this way a spectral equivalence at the level of the QES sectors is established.  
For some models a complete spectral equivalence can also be established \cite{DHL}.

In the next section we make the explicit connection between the above results and the algebraic QES part of the AQRM. 
This establishes a spectral equivalence between the QES eigenvalues of the AQRM hamiltonian and the QES sector of the 
Schr\"odinger operator. 
This equivalence is then extended to a complete spectral equivalence between the two systems, 
thus providing a generalisation of the known mapping \cite{Koc} between the quantum Rabi model and the QES P\"oschl-Teller potential. 
The paper concludes with a brief discussion of the results and their implications.

\section{Results}

We begin by collecting the relevant results \cite{LB1} for the AQRM. 

\subsection{Algebraic equations for the asymmetric quantum Rabi model}

By making use of the Bargmann realisation \cite{Schw}
\begin{equation}
a^\dagger \to z ,  \quad a \to \frac{d}{dz}
\end{equation}
the hamiltonian (\ref{ham}) is transformed to 
\begin{equation}
H= \Delta \, \sigma_z+\epsilon\,\sigma_x + \omega \, z \frac{d}{dz} + g \, \sigma_x \left(z+\frac{d}{dz} \right) \,.
\label{hamB}
\end{equation}
It then follows that in terms of the two-component wavefunction  
\begin{equation}
\psi(z) = \left( \begin{array}{c}
\psi_+(z) \\
\psi_-(z) \end{array} \right) , 
\end{equation}
the Schr\"odinger equation $H \psi = E \psi$ gives rise to the pair of coupled equations 
\begin{eqnarray}
(\omega z +g) \frac{{d\psi_+}}{{dz}} + (g z + \epsilon- E) \psi_+ + \Delta \psi_- &=& 0 \,,\\
(\omega z -g) \frac{{d\psi_-}}{{dz}} - (g z + \epsilon+ E) \psi_- + \Delta \psi_+ &=& 0 \,.
\end{eqnarray}

Two sets of solutions for the components $\psi_+(z)$ and $\psi_-(z)$ have been obtained. 
For the first set, the substitution $\psi^1_{\pm} (z)= {\mathrm e}^{-gz/\omega} \phi^1_\pm(z)$ leads to the coupled equations
\begin{eqnarray}
\left[(\omega z +g) \frac{{d}}{{dz}} - \left(\frac{g^2}{\omega} + E - \epsilon\right) \right] \phi^1_+(z)  = - \Delta \phi^1_-(z) \,,\label{phi}\\
\left[(\omega z -g) \frac{{d}}{{dz}} - \left(2gz -\frac{g^2}{\omega} + E + \epsilon\right) \right] \phi^1_-(z)  = - \Delta \phi^1_+(z)  \,.
\end{eqnarray}
Eliminating $\phi^1_-(z)$ gives the second order differential equation
\begin{eqnarray}
(\omega z - g) (\omega z +g) \frac{{d^2\phi^1_+(z)}}{{d^2z}} & \nonumber\\
+ \left[ - 2g\omega z^2 + (\omega^2 - 2 g^2 - 2 E \omega) z + 
\frac{g}{\omega} (2g^2 - \omega^2 - 2 \epsilon\omega)\right] \frac{{d\phi^1_+(z)}}{{dz}} & \nonumber\\
+\left[  2g\left(  \frac{g^2}{\omega}  + E -\epsilon\right) z + E^2 - \Delta^2 - \epsilon^2 + \frac{2\epsilon g^2}{\omega} - 
\frac{g^4}{\omega^2}     \right] \phi^1_+(z) = 0 \,.
\label{ode1}
\end{eqnarray}
For the algebraic QES part of the eigenspectrum the wavefunction component is given in the factorised form
\begin{equation}
\psi^1_+ (z)= {\mathrm e}^{-gz/\omega} \prod_{i=1}^n (z-z_i) \,, 
\label{fac}
\end{equation}
where the  $z_i$ satisfy the set of algebraic equations 
\begin{equation}
\sum_{j \ne i}^n \frac{2\omega}{z_i - z_j} =  \frac{n \omega^2 + 2 \epsilon \omega}{\omega z_i - g} + \frac{n \omega^2 -\omega^2}{\omega z_i + g} + 2g
\label{alg1}
\end{equation}
for $i=1,\ldots,n$. 
The system parameters obey the constraint 
\begin{equation}
\Delta^2 + 2 n g^2  + 2 \omega g \sum_{i=1}^n z_i = 0 \,. \label{con1}
\end{equation}
The energy of these states is given by 
\begin{equation}
E= n \omega  - \frac{g^2}{\omega} + \epsilon \,.  \label{energy+}
\end{equation}
The corresponding wavefunction component $\psi^1_- (z)$ is determined using the result (\ref{fac}) and equation (\ref{phi}).

The other set of solutions follow from the 
substitution $\psi^2_{\pm} (z)= {\mathrm e}^{gz/\omega} \phi^2_\pm(z)$,  leading to the coupled equations
\begin{eqnarray}
\left[(\omega z +g) \frac{{d}}{{dz}} + \left(2gz +\frac{g^2}{\omega} - E + \epsilon\right) \right] \phi^2_+(z)  = - \Delta \phi^2_-(z) \,, \\
\left[(\omega z -g) \frac{{d}}{{dz}} - \left(\frac{g^2}{\omega} + E + \epsilon\right) \right] \phi^2_-(z)  = - \Delta \phi^2_+(z) \,. \label{phi2}
\end{eqnarray}
Eliminating $\phi^2_+(z)$ gives the second order differential equation
\begin{eqnarray}
(\omega z - g) (\omega z +g) \frac{{d^2\phi^2_-(z)}}{{d^2z}} & \nonumber\\
+ \left[ 2g\omega z^2 + (\omega^2 - 2 g^2 - 2 E \omega) z - 
\frac{g}{\omega} (2g^2 - \omega^2 + 2 \epsilon\omega)\right] \frac{{d\phi^2_-(z)}}{{dz}} & \nonumber\\
+\left[  -2g\left(  \frac{g^2}{\omega}  + E +\epsilon\right) z + E^2 - \Delta^2 - \epsilon^2 - \frac{2\epsilon g^2}{\omega} - 
\frac{g^4}{\omega^2}     \right] \phi^2_-(z) = 0 \,.
\label{ode2}
\end{eqnarray}
For the QES component of the eigenspectrum these equations are solved for the wavefunction components  in the form 
\begin{equation}
\psi^2_- (z)= {\mathrm e}^{gz/\omega} \prod_{i=1}^n -(z-z_i) \label{fac2}
\end{equation}
where the roots $\{z_k\}$ satisfy the algebraic equations
\begin{equation}
\sum_{j \ne i}^n \frac{2\omega}{z_i - z_j} = \frac{n \omega^2 -\omega^2}{\omega z_i - g}  + \frac{n \omega^2 - 2 \epsilon \omega}{\omega z_i + g} - 2g
\label{alg2}
\end{equation}
for $i=1,\ldots,n$. 
The system parameters obey the constraint 
\begin{equation}
\Delta^2 + 2 n g^2  - 2 \omega g \sum_{i=1}^n z_i = 0 \,,  \label{con2}
\end{equation}
with energy  
\begin{equation}
E= n \omega  - \frac{g^2}{\omega} - \epsilon \,.  \label{energy-}
\end{equation}
The wavefunction component $\psi^2_+ (z) = {\mathrm e}^{gz/\omega} \phi^2_+(z)$ 
follows from (\ref{fac2}) and  (\ref{phi2}).

The symmetry between the two sets of solutions has been noted \cite{Heun2,LB1}.
Namely the algebraic equations (\ref{alg1}) and (\ref{alg2}) are equivalent under the 
transformation $z_i \leftrightarrow -z_i$, $\epsilon \leftrightarrow - \epsilon$. 
This corresponds to the related symmetry $\psi_+^1(z,\epsilon) = \psi_-^2(-z,-\epsilon)$, $\psi_-^1(z,\epsilon) = \psi_+^2(-z,-\epsilon)$
in the wavefunction components.
The $-$ sign appears in equation (\ref{fac2}) to ensure this symmetry.

\subsection{Constraint polynomials}

The constraint polynomials $P_n(x,y)$ for the AQRM were defined in \cite{LB1} following the work of K\'us \cite{Kus} 
on the (symmetric) quantum Rabi model. 
These polynomials were derived in the framework of finite-dimensional irreducible representations of $\mathfrak{sl}_2$ in the confluent Heun picture of the AQRM \cite{Wakayama1}. 
The polynomials $P_k(x,y)$ of degree $k$ are defined via the three-term recursion relation \cite{LB1,Wakayama1,Wakayama2}
\begin{eqnarray}
P_k(x,y) &=& \left[ k x + y - k^2 \omega^2 - 2k \epsilon \, \omega \right] P_{k-1}(x,y) \nonumber \\ 
&& - k(k-1)(n-k+1) x \omega^2 P_{k-2}(x,y)
 \,, \label{rec1}
\end{eqnarray}
with $P_0(x,y) = 1$ and $P_1(x,y) = x + y - \omega^2 - 2 \epsilon \omega$.
The zeros of the constraint polynomials, 
\begin{equation}
P_n((2g)^2,\Delta^2) = 0,
\end{equation}
define the QES, or Juddian solutions of the model, with in this case the energy given by ({\ref{energy+}).
Although the precise connection is not at all obvious, the constraint polynomials are also of the form (\ref{con1}) and (\ref{con2}) 
in terms of the Bethe ansatz roots $\{z_k\}$.
We will touch on this point further below.

Having laid out the relevant results for the AQRM we  are now ready to make the connection with the Schr\"odinger equation (\ref{SE}).

\subsection{Equivalent QES Schr\"odinger potentials}

Beginning with the algebraic equations, comparison of (\ref{alg1}) and (\ref{con1}) with (\ref{BAE}) and (\ref{calE}) gives
\begin{equation}
A_+ = - 2g/\omega, \,\, B_+ = n + 2  \epsilon/\omega, \,\, C_+ =  n -1, \,\,  \gamma = 2  g/\omega, 
\label{c1}
\end{equation}
and
\begin{equation}
{\cal E} = - {\Delta^2}/{\omega^2} - 2 n \,{g^2}/{\omega^2}. 
\label{Eres}
\end{equation}
Likewise, comparison of (\ref{alg2}) and (\ref{con2}) with (\ref{BAE}) and (\ref{calE}) gives
\begin{equation}
A_- = 2g/\omega, \,\, B_- = n-1, \,\, C_- = n - 2  \epsilon/\omega, \,\, \gamma = 2  g/\omega, 
\label{c2}
\end{equation}
with ${\cal E}$ also given by (\ref{Eres}). 
In each case $M=n$ and we identify $v_j = - z_j$.

For each case the corresponding wavefunction $\psi(x)$ is given by (\ref{psi}) with the Schr\"odinger potential $V(x; A,B,C,\gamma)$ given by (\ref{pot}).
This establishes the spectral equivalence between the QES sector of the AQRM on the one hand, and a QES hyperbolic Schr\"odinger potential on the other. 
For the set of parameters (\ref{c1}) the Schr\"odinger potential  is 
\begin{eqnarray}
V_+(x) &=& \frac{\epsilon^2}{\omega^2} + \frac{g^4}{\omega^4} \sinh^2 x +  \frac{g^2}{\omega^2} (1- \cosh x) + \frac{2 g^2 \epsilon}{\omega^3}   (1 + \cosh x)  
\nonumber\\ 
& & - \frac{4n^2-1}{8(\cosh x +1)} + \frac{(2n+1 + 4\epsilon/\omega)(2n+3 + 4\epsilon/\omega)}{8(\cosh x -1)}.
\label{pot1}
\end{eqnarray} 
The parameters (\ref{c2}) give
\begin{eqnarray}
V_-(x) &=& \frac{\epsilon^2}{\omega^2} + \frac{g^4}{\omega^4} \sinh^2 x +  \frac{g^2}{\omega^2} (1 + \cosh x) - \frac{2 g^2 \epsilon}{\omega^3}   (1 - \cosh x)  
\nonumber\\ 
& & + \frac{4n^2-1}{8(\cosh x -1)} - \frac{(2n+1 - 4\epsilon/\omega)(2n+3 - 4\epsilon/\omega)}{8(\cosh x +1)}.
\label{pot2}
\end{eqnarray}

We will demonstrate that these hyperbolic Schr\"odinger potentials are in fact generalised QES P\"oschl-Teller potentials.   
Consider first the potential (\ref{pot1}), which can be rewritten as
\begin{eqnarray}
V_+(x) &=& \frac{\epsilon^2}{\omega^2} + \frac{g^2}{\omega^2}\left(1+\frac{2\epsilon}{\omega}\right) - \frac{g^2}{\omega^2}\left(1-\frac{2\epsilon}{\omega}\right) \cosh x +
\frac{g^4}{\omega^4} \sinh^2 x 
\nonumber\\ 
& & + \case14 \left( (2n+1)^2 + 8 \epsilon/\omega \, (1+n + \epsilon/\omega)  \right) {\mathrm{csch}}^2 x 
\nonumber\\ 
& & + \case12 \left( 2n+1 + 4 \epsilon/\omega \, (1+n + \epsilon/\omega) \right) \coth x \, {\mathrm{csch}} \,x . 
\label{QESV1}
\end{eqnarray} 
Similarly the potential (\ref{pot2}) is 
\begin{eqnarray}
V_-(x) &=& \frac{\epsilon^2}{\omega^2} + \frac{g^2}{\omega^2}\left(1-\frac{2\epsilon}{\omega}\right) + \frac{g^2}{\omega^2}\left(1+\frac{2\epsilon}{\omega}\right) \cosh x +
\frac{g^4}{\omega^4} \sinh^2 x 
\nonumber\\ 
& & + \case14 \left( (2n+1)^2 - 8 \epsilon/\omega \, (1+n - \epsilon/\omega)  \right) {\mathrm{csch}}^2 x 
\nonumber\\ 
& & - \case12 \left( 2n+1 - 4 \epsilon/\omega \, (1+n - \epsilon/\omega) \right) \coth x \, {\mathrm{csch}} \,x .
\label{QESV2}
\end{eqnarray} 
The hyperbolic functions appearing in these potentials are precisely those given in equation (5.11) of reference \cite{GKO}, therein corresponding  to 
canonical form IIb (Case 2a).
Here the various constant terms and prefactors are essential to the spectral equivalence with the AQRM.
In principle the constant terms in $V_\pm(x)$ could be absorbed into the energy (\ref{Eres}).
We also note that, as pointed out in \cite{GKO}, the exactly-solvable case is when $g=0$, 
here corresponding to the absence of the light-matter interaction term in the AQRM, as we should expect.

The corresponding wavefunctions are, respectively,   
\begin{eqnarray}
\Psi_+(x) &=& \frac{\exp \left( -\frac{g^2}{\omega^2} \cosh x \right) }
{(\cosh x -1)^{\frac14(2n+1+4\epsilon/\omega)}(\cosh x+1)^{\frac14(2n-1)}} \prod_{j=1}^n \left(\frac{g}{\omega} \cosh x + v_j \right) ,
\nonumber \\
&& \\
\Psi_-(x) &=& \frac{\exp \left( \frac{g^2}{\omega^2} \cosh x \right)}
{(\cosh x -1)^{\frac14(2n-1)}(\cosh x+1)^{\frac14(2n+1-4\epsilon/\omega)}} \prod_{j=1}^n \left(\frac{g}{\omega} \cosh x + v_j \right). 
\nonumber\\
&&   
\end{eqnarray}

As a concrete example, consider $n=1$. 
The algebraic Bethe ansatz equations (\ref{BAE}) reduce to
\begin{equation}
	A_+ +\frac{B_+}{v_1+\gamma/2}+\frac{C_+}{v_1-\gamma/2}=0.
\end{equation}
For the parameter set (\ref{c1}), the solution is 
\begin{equation}
	v_1=\frac{\omega ^2-2 g^2+2  \epsilon \omega  }{2 g \omega }.
\end{equation}
The energy follows as  
\begin{equation}
	{\cal E}= - 1 + 2 g^2/\omega^2 - 2 \epsilon/\omega    . 
\end{equation}
It should be noted that this energy is equivalent to the general result (\ref{Eres}) due to the $n=1$ constraint relation 
\begin{equation}
\Delta^2 + 4g^2 - \omega^2 - 2\epsilon \omega =0.
\end{equation}
The corresponding results for the wavefunction and potential are 
\begin{eqnarray}
	\Psi_+(x)&=&(\cosh x-1)^{-3/4-\epsilon/\omega}(\cosh x+1)^{-1/4}\exp \left( -\frac{g^2}{\omega^2}\cosh x\right) \nonumber\\
	&&\times\left(\frac{g}{\omega}\cosh x+\frac{\omega^2-2g^2+2\epsilon\omega}{2g\omega}\right),
\end{eqnarray}

\begin{eqnarray}
	V_+(x)&=&\frac{\epsilon^2}{\omega^2}+\frac{g^4}{\omega^4}\sinh^2 x + \frac{g^2}{\omega^2}(1-\cosh x)+\frac{2g^2\epsilon}{\omega^3}(1+\cosh x)  \nonumber\\
	&&-\frac{3}{8(\cosh x+1)}+\frac{(3+4\epsilon/\omega)(5+4\epsilon/\omega)}{8(\cosh x-1)}. 
\end{eqnarray}
It can be readily verified that ${\cal E},\Psi(x)$ and $V(x)$ satisfy the Schr\"odinger equation (\ref{SE}).

Similarly the parameter set (\ref{c2}) leads to the solution 
\begin{equation}
v_1 = \frac{2 g^2-\omega ^2+2 \epsilon \omega   }{2 g \omega }.
\end{equation}
and thus the energy
\begin{equation}
	{\cal E}= - 1 + 2 g^2/\omega^2 + 2 \epsilon/\omega    . 
\end{equation}
Here the corresponding constraint relation 
\begin{equation}
\Delta^2 + 4g^2 - \omega^2 + 2\epsilon \omega =0,
\end{equation}
ensures (\ref{Eres}) is satisfied.
In this case the Schr\"odinger equation (\ref{SE}) is satisfied with the wavefunction and potential 
\begin{eqnarray}
\Psi_-(x)&=&(\cosh x-1)^{-1/4}(\cosh x+1)^{-3/4+\epsilon/\omega}\exp \left( \frac{g^2}{\omega^2}\cosh x\right)  \nonumber\\
&&\times\left(\frac{g}{\omega}\cosh x+\frac{2g^2+2\epsilon\omega-\omega^2}{2g\omega}\right),
\end{eqnarray}
\begin{eqnarray}
V_-(x)&=&\frac{\epsilon^2}{\omega^2}+\frac{g^4}{\omega^4}\sinh^2 x + \frac{g^2}{\omega^2}(1+\cosh x)-\frac{2g^2\epsilon}{\omega^3}(1-\cosh x)  \nonumber\\
&&+\frac{3}{8(\cosh x-1)}-\frac{(3-4\epsilon/\omega)(5-4\epsilon/\omega)}{8(\cosh x+1)}. 
\end{eqnarray}

To illustrate the QES spectral equivalence between the two systems more generally, consider the 
eigenspectrum of the AQRM shown in Figure \ref{rabi} as a function of the coupling $g$ at the particular 
asymmetry value $\epsilon=0.3$. 
The QES points are indicated as circles. 
The corresponding energy values of the QES generalised P\"oschl-Teller potentials are shown in Figure \ref{pt}.

\begin{figure}[t]
\begin{center}
\includegraphics[width=1.0\columnwidth]{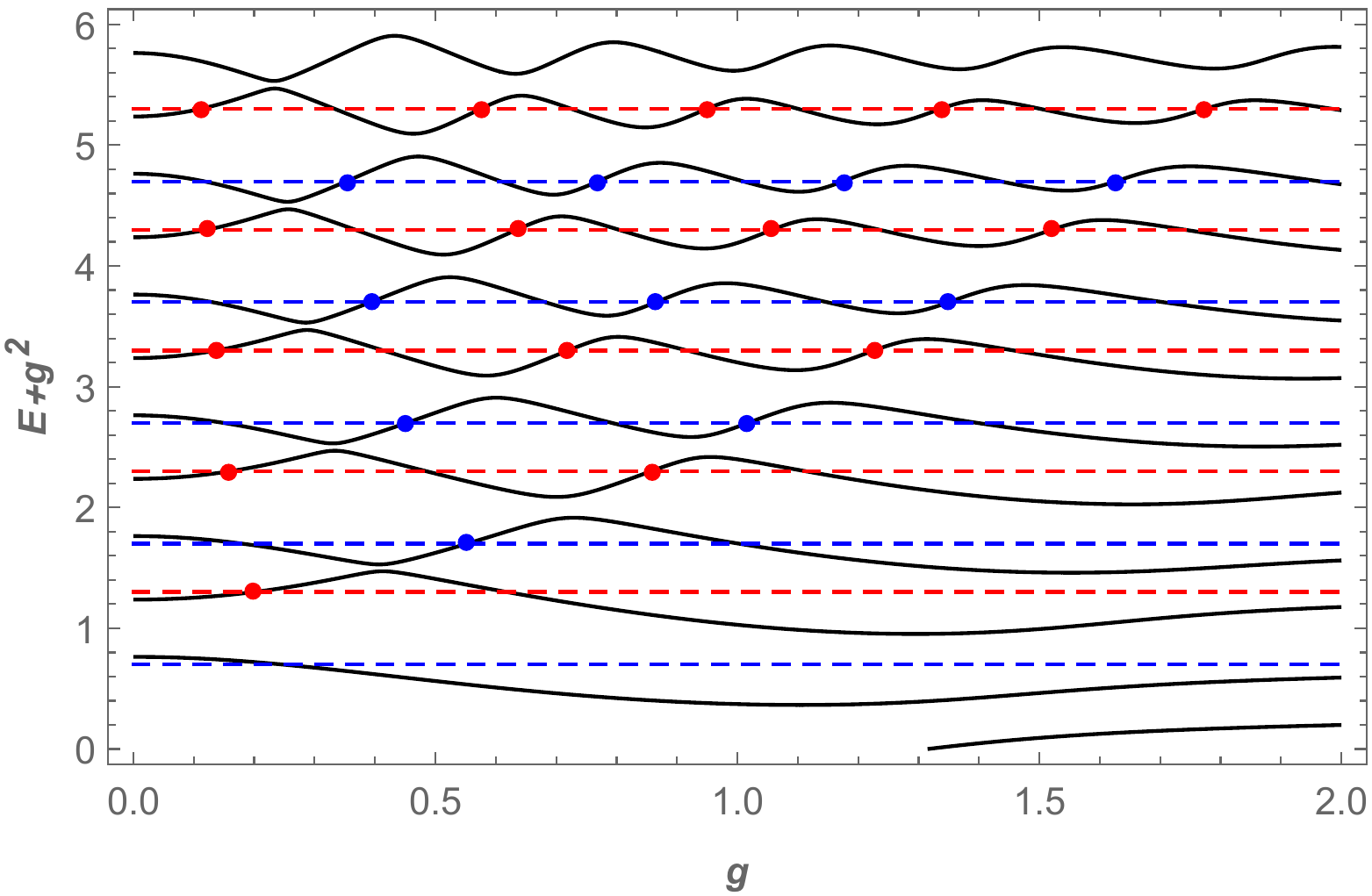}
\caption{Rescaled lowest energy levels $E+g^2$ in the eigenspectrum of the AQRM (\ref{ham}) as a function of the light-matter coupling $g$. 
The parameter values are $\Delta = 1.2$, $\omega = 1$ and $\epsilon = 0.3$. 
The blue lines are the energy $E+g^2=n  -\epsilon$ for $n=1,2,3,4,5$. 
The red lines are the energy $E+g^2=n +\epsilon$ for $n=1,2,3,4,5$.
In each case the circles indicate the QES exceptional points. 
For the given parameter values there are $n$ QES points on the red lines and $n-1$ QES points on the blue lines. 
The precise values of $g$ at the QES points can be determined from the roots of the constraint polynomials.
As $\epsilon \to 0$ the QES points become doubly degenerate crossing points.
The energy levels have been obtained using Braak's $G$-function  \cite{Braak}.}
\label{rabi}
\end{center}
\end{figure}

\begin{figure}[t]
\begin{center}
\includegraphics[width=1.0\columnwidth]{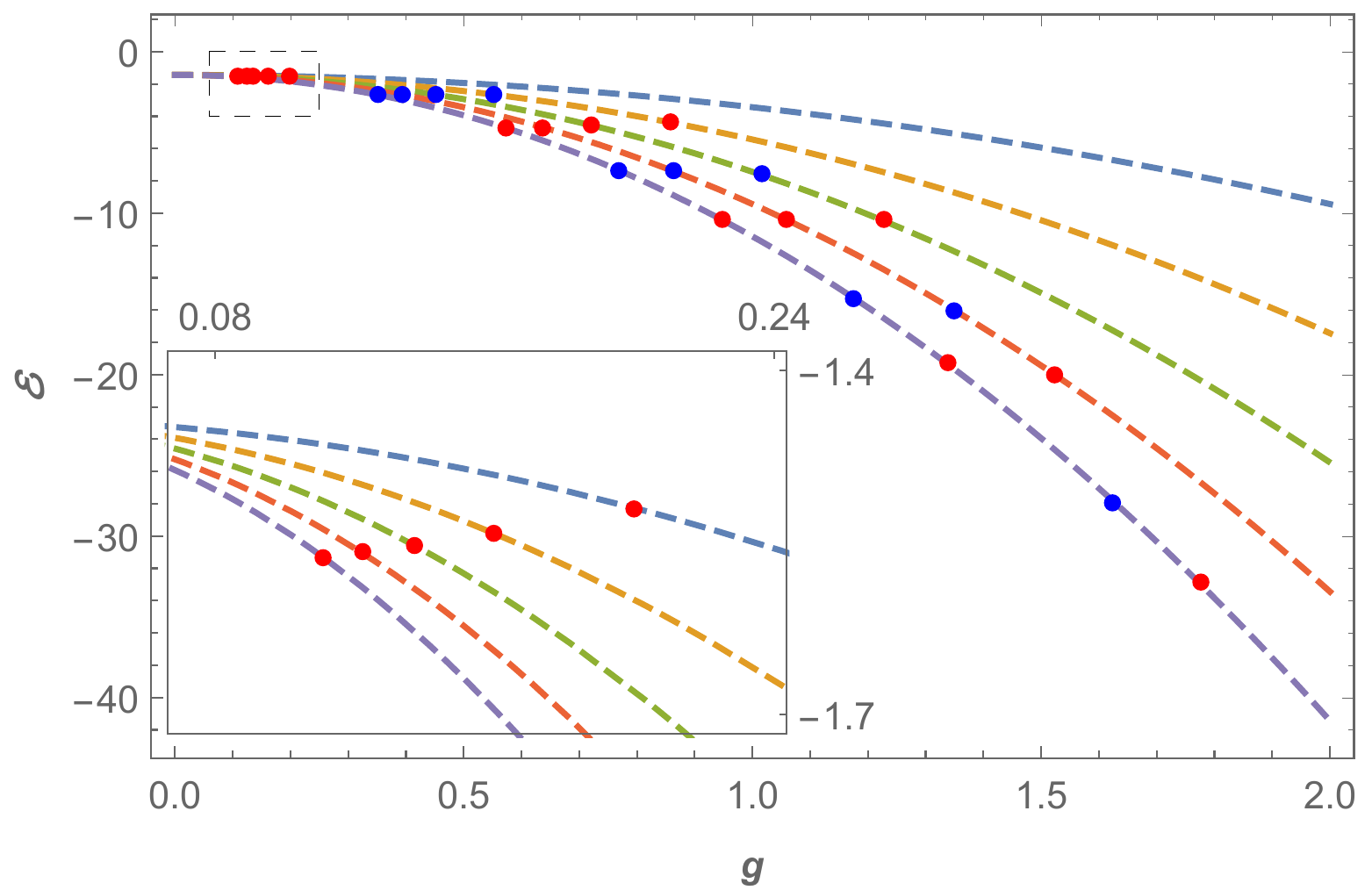}
\caption{The energy levels ${\cal E}$ (\ref{Eres}) of the QES generalised P\"oschl-Teller potentials (\ref{pot1}) and (\ref{pot2}) 
with parameter values $\Delta = 1.2$, $\omega = 1$ and $\epsilon = 0.3$.
Here, from top to bottom, $n=1,2,3,4,5$. 
The QES points are indicated by circles. 
%
%
The inset shows a magnification of the indicated region.
The QES points are spectral equivalent to the QES points indicated in Figure \ref{rabi}.
}
\label{pt}
\end{center}
\end{figure}

\subsection{QES general form and constraint polynomials}

A second order differential equation has a QES sector if it can be written in the form 
\begin{eqnarray}
P ( z ) {\frac {{\rm d}^{2} y(z)}{{\rm d}{z}^2}}
  &+& \left[ Q \left( z \right) -\frac{ n-1 }{2} P'( z ) \right] \frac{ {\rm d}{y(z)} }{ {\rm d} z} \nonumber\\
  &+& \left[ R  -\frac{n}{2} Q' ( z ) +\frac{n ( n-1 ) }{12} P'' ( z )  \right] { y} \left( z \right) =0 ,
\end{eqnarray}
where in general $P(z)$ is a quartic polynomial, $Q(z)$ is a quadratic polynomial, $R$ is a constant and $n$ is a non-negative integer \cite{GKO}. 
Comparing this form with equation (\ref{ode1}) for the $\phi_{1}^+(z)$ component,  the polynomials are thus 
\begin{eqnarray}
P(z)&=& {\omega}^{2}{z}^{2}-{g}^{2}  ,\\
Q(z)&=& - 2 g \omega \, {z}^{2} - \left( n {\omega}^{2}+2 {\epsilon} \omega \right) z - 
\frac {g}{\omega} \left(  {\omega}^{ 2} + 2 {\epsilon} \omega -2 {g}^{2} \right) ,\\
\phantom{(z)}R &=& \frac{1}{3} {n}^{2} {\omega}^{2}+\frac{1}{6} n {\omega} ^{2}+n {\epsilon} \omega-2 n {g}^{2} -{\Delta}^{2} ,
\end{eqnarray}
along with the energy relation (\ref{energy+}).
The algebraic sector has $n+1$ eigenfunctions of the form (\ref{fac}), 
%
%
one of which is the $\Delta=0$ case with $z_i= -g/\omega$, 
corresponding to the degenerate atomic limit in the Rabi model \cite{LB1}.

The polynomials for $\phi_{2}^-(z)$ satisfying equation (\ref{ode2}) are
\begin{eqnarray}
P(z)&=& {\omega}^{2}{z}^{2}-{g}^{2}  ,\\
Q(z)&=& 2 g \omega \, {z}^{2} -( n {\omega}^{2}-2 {\epsilon} \omega ) z + 
\frac {g}{\omega} \left(  {\omega}^{ 2} - 2 {\epsilon} \omega -2\,{g}^{2} \right)  ,\\
\phantom{(z)}R &=& \frac{1}{3} {n}^{2} {\omega}^{2}+\frac{1}{6} n {\omega} ^{2}-n {\epsilon} \omega-2 n {g}^{2} -{\Delta}^{2} ,
\end{eqnarray}
along with the energy relation (\ref{energy-}).

Here the particular polynomial $P(z)$ 
corresponds to canonical form IIb in the classification of QES spectral problems \cite{GKO}, 
discussed therein as Case 2a and Case 2b depending on the domain $z$.
The corresponding change of variables is as given in (\ref{trans}) below.
Within the general QES formalism contact can also be made with the three-term recursion relations defining the constraint polynomials \cite{FGR}.
In this way the known constraint polynomials for the AQRM can be recovered.
These same constraint polynomials appear in the solutions involving the generalised P\"oschl-Teller potentials.

The polynomials $P_k(x,y)$ arise from a  generating function type of solution to the confluent Heun picture of the AQRM \cite{Wakayama2}.  
Relations (\ref{fac}) and (\ref{alg1}) arise from assuming a product type of solution 
\begin{equation}
\phi_{+}^1(z)=\prod_{i=1}^n (z-z_i), 
\end{equation}
which we insert into (\ref{ode1}) and equate coefficients of powers of $z$. 
The coefficient at order $z^{n+1}$ is zero if $E$ satisfies (\ref{energy+}), 
the constraint (\ref{con1}) is the coefficient of  $z^{n}$ and the 
coefficients of lower order powers of $z$ specify the Bethe ansatz equations (\ref{alg1}).
If we seek a solution to (\ref{ode1}) in the form of a generating function, 
\begin{equation}
\phi_{+}^1(z)=\sum_{k=0}^\infty R_k(n,\epsilon,\omega,\Delta)\, z^k, 
\label{C2}
\end{equation}
we find a 4-term recursion relation for the coefficients $R_k$,
and cannot easily deduce $R_{n+k}= 0$ for $ k=1,2,\dots.$ 
However, one further variable transformation will allow a direction connection to be made between the approach discussed in \S 2.1 and the constraint polynomials $P_n(x,y)$.
Setting $E$ as per equation (\ref{energy+}) and applying the variable changes 
\begin{equation}
z=-\frac{g}{\omega} \frac{ u+1}{u-1}, 
\quad\quad 
y(u) =\omega(u-1)^{-n} f(u), 
\label{vtf} 
\end{equation}
to equation (\ref{ode1}) gives 
\begin{eqnarray}
&& u(u-1)^2 \omega^2 \frac{d^2 f(u)}{du^2}  \nonumber\\
&& + 
\left( (1-n)\omega^2 u^2 + (2\epsilon\omega +(2n-1)\omega^2  - 4g^2 )u - \omega(2\epsilon +n\omega) \right) \frac{df(u)}{du} \nonumber\\
&& - \Delta^2 f(u) =0.
\label{ode4}
\end{eqnarray}
Note that a further variable change maps this equation to the confluent Heun equation of relevance to the AQRM \cite{Heun2,others2,Wakayama1}.
Here we work with the form (\ref{ode4}) because it explicitly includes the special solutions $\Delta^2=0$, as we note below.

The function
\begin{equation}
f(u)=\sum_{k=0}^\infty Q_k(n,\epsilon,\omega,\Delta) \,u^k
\label{bd}
\end{equation}
 is a solution of (\ref{ode4}) provided the coefficients $Q_k := Q_k(n,\epsilon,\omega,\Delta)$ satisfy the three-term recurrence relation 
\begin{eqnarray}
&& \omega(k+1)(2\epsilon+n\omega-k\omega) Q_{k+1}  \nonumber\\
&& \qquad = -Q_k(\omega^2( 2k^2 -2kn-k) - 2k\epsilon\omega + 4kg^2 +\Delta^2) \nonumber\\
&& \qquad \quad \, +Q_{k-1} (1-k) \omega^2 (n-k+1),
\label{recur}
\end{eqnarray}
with initial condition $Q_{-1}=0$ and $Q_0$. 
When $k=n+1$ 
\begin{equation}
\omega(n+2)(2\epsilon-\omega)\, Q_{n+2} =   \left( ( 2\,\epsilon\,\omega-
4\,{g}^{2}-\omega^2)(n+1) -{\Delta}^{2} \right) Q_{n+1}.
\end{equation}
Setting 
\begin{equation}
Q_{n+1}(n,\epsilon, \omega,\Delta)=0
\label{qc}
\end{equation}
leads to  $Q_{n+1+k}=0 $ for $k =0,1,\dots$ and the  series (\ref{bd})  truncates to a polynomial.  
More explicitly,
$Q_{n+1}=0$ sets the coefficient of $u^n$ in
({\ref{C2}) to zero when $f(u)$ takes the form (\ref{vtf}), with the coefficients of lower order terms in the expansion defining  
$Q_k(n,\epsilon,\omega,\Delta)$ in terms of $Q_n(n,\epsilon,\omega,\Delta)$,  
resulting in the QES solutions $f(u)$ of the ARQM model.

The connection between $Q_{n+1}(n,\epsilon,\omega,\Delta)$ and  
the constraint polynomial  $P_n((2g)^2,\Delta^2)$ is 
\begin{equation}
Q_{n+1}(n,\epsilon,\omega,\Delta) = \frac{(-1)^{n+1} \Delta^2} {\omega^{n+1} 2^{n+1} (n+1)! \prod_{k=0}^n (\epsilon+k\omega/2) }  
P_n((2g)^2,\Delta^2).
\label{qp}
\end{equation}
The constraint  
$P_n((2g)^2,\Delta^2)=0$ does not include the degenerate atomic limit solutions
of the AQRM that arise when $\Delta^2=0$. These solutions are built into
(\ref{qc}) as can be deduced from the factor $\Delta^2$ on the right-hand side of (\ref{qp}).

We also note that though 
the polynomials $Q_k$ satisfy a 3-term recurrence relation, they are 
not  orthogonal polynomials in the usual sense and are instead said to be weakly orthogonal \cite{FGR}. 

The variable transformations (\ref{vtf}) can be unravelled to find the relation between the polynomials $Q_k$ and the Bethe ansatz roots $\{z_k\}$.
We have  
\begin{equation}
\frac{(-1)^n}{\omega} \prod_{k=1}^{n} \left[ \left( \frac{g}{\omega} +z_k \right) u + 
\left( \frac{g}{\omega} -z_k\right) \right]
  = \sum_{k=0}^{n} Q_k \, u^k, 
\end{equation}
with 
\begin{equation}
Q_0= \frac{(-1)^n }{\omega^{n}} \prod_{k=1}^n \left(\frac{g}{\omega}- z_k\right)\,.
\end{equation}
Expanding the left-hand side,  the polynomials $Q_k$
are  expressed in terms of the Bethe ansatz roots $\{z_k\}$ via  
\begin{equation}
Q_k=
\frac{(-1)^n}{\omega}S_{n-k} \left (\frac{g -\omega z_1} { g +\omega z_1},
\frac{g -\omega z_2} { g +\omega z_2}   ,\dots,     
\frac{g -\omega z_n} { g +\omega z_n} \right) \, 
 \prod_{k=1}^n  \left( \frac{g}{\omega} +z_k\right) , 
 \label{Qresult}
\end{equation}
where $S_j(x_1,\dots,x_n)$ is the $j^{\rm th}$ symmetric polynomial on $n$ variables.

This argument can similarly be repeated for the other QES sector of the AQRM by considering the equation satisfied by  $\phi^{2}_{-}(z)$.

\subsection{Complete spectral equivalence}

So far we have demonstrated the spectral equivalence between the QES energies of the AQRM on the one hand,  
and hyperbolic Schr\"odinger potentials on the other.
In fact this spectral equivalence is {\it complete}.
This can be shown by applying the change of variable
\begin{equation}
z = \frac{g}{\omega} \cosh x 
\label{trans}
\end{equation}
in the second order differential equations (\ref{ode1}) and (\ref{ode2}), along with the transformations
\begin{eqnarray}
\phi_+^1(x) &=&
\left( \cosh x  -1 \right)^{{\frac {2 E\omega+2{\epsilon} \omega+2{g}^{2}+{\omega}^{2}}{4{\omega}^{2}}}} 
\left( \cosh x  +1 \right) ^{{\frac {2 E\omega-2{\epsilon} \omega+2{g}^{2}-{\omega}^{2}}{4{\omega}^{2}}}} \nonumber\\
&& \times  \exp \left( {\frac{g^2}{\omega^2} \cosh x }\right) \, \Psi_+(x)  , \label{wav1}\\
\phi_-^2(x) &=&\left( \cosh x  -1 \right) ^{{\frac {2 E \omega+2 \epsilon \omega+2 {g}^{2}-{\omega}^{2}}{4{\omega}^{2}}}} 
\left( \cosh x  +1 \right) ^{{\frac {2 E \omega-2 {\epsilon} \omega+ 2{g}^{2}+{\omega}^{2}}{4{\omega}^{2}}}} \nonumber\\
&& \times \exp \left( -{\frac{g^2}{\omega^2} \cosh x }\right) \, \Psi_-(x) .  \label{wav2}
\end{eqnarray}
The differential equations for $\phi_+^1(x)$ and $\phi_-^2(x)$ then transform to Schr\"odinger equations of the form (\ref{SE}), with 
wavefunctions $\Psi_\pm(x)$ and hyperbolic potentials 
\begin{eqnarray}
V_+(x) &=&  \frac{\epsilon^2}{\omega^2} + \frac{g^4}{\omega^4} \sinh^2 x +  \frac{g^2}{\omega^2} (1- \cosh x) + \frac{2 g^2 \epsilon}{\omega^3}   (1 + \cosh x) 
\nonumber\\ 
& & + \frac{(2E \omega + 2 \epsilon \omega + 2g^2 + 3 \omega^2)(2E \omega + 2 \epsilon \omega + 2g^2 +  \omega^2)}{8\omega^4 (\cosh x -1)} \nonumber\\ 
&& - \frac{(2E \omega - 2 \epsilon \omega + 2g^2 + \omega^2)(2E \omega - 2 \epsilon \omega + 2g^2 -  \omega^2)}{8\omega^4(\cosh x +1)}, 
\label{full1} \\
V_-(x) &=&  \frac{\epsilon^2}{\omega^2} + \frac{g^4}{\omega^4} \sinh^2 x +  \frac{g^2}{\omega^2} (1 + \cosh x) - \frac{2 g^2 \epsilon}{\omega^3}   (1 - \cosh x)
\nonumber\\ 
& & + \frac{(2E \omega + 2 \epsilon \omega + 2g^2 +  \omega^2)(2E \omega + 2 \epsilon \omega + 2g^2 -  \omega^2)}{8\omega^4 (\cosh x -1)} \nonumber\\ 
&& - \frac{(2E \omega - 2 \epsilon \omega + 2g^2 + 3 \omega^2)(2E \omega - 2 \epsilon \omega + 2g^2 +  \omega^2)}{8\omega^4(\cosh x +1)}. 
\label{full2}
\end{eqnarray} 
The corresponding energy is given by 
\begin{equation}
{\cal E_\pm} = - 2 E g^2/\omega^3 - 2g^4/\omega^4 - {\Delta^2}/{\omega^2} \pm 2 g^2 \epsilon/\omega^3 . 
\label{Ecom}
\end{equation}

The potentials (\ref{full1}) and (\ref{full2}) can be simplified to some extent.
We write them in the form
\begin{eqnarray}
V_+(x) &=&  \frac{\epsilon^2}{\omega^2} + \frac{g^2}{\omega^2}\left(1+\frac{2\epsilon}{\omega}\right) - \frac{g^2}{\omega^2}\left(1-\frac{2\epsilon}{\omega}\right) \cosh x +
\frac{g^4}{\omega^4} \sinh^2 x 
\nonumber\\ 
& & + \left[ \left(E + {g^2}/{\omega} + \omega/2 \right)^2 +  \epsilon^2 + \epsilon \omega \right]  {\mathrm{csch}}^2 x  \nonumber\\ 
&& + (2\epsilon/\omega+1) (E \omega + g^2 + \omega^2/2) \coth x \, {\mathrm{csch}} \,x , 
\label{full1} \\
V_-(x) &=& \frac{\epsilon^2}{\omega^2} + \frac{g^2}{\omega^2}\left(1-\frac{2\epsilon}{\omega}\right) + \frac{g^2}{\omega^2}\left(1+\frac{2\epsilon}{\omega}\right) \cosh x +
\frac{g^4}{\omega^4} \sinh^2 x  
\nonumber\\ 
& & + \left[ \left(E + {g^2}/{\omega} + \omega/2 \right)^2 +  \epsilon^2 - \epsilon \omega \right]  {\mathrm{csch}}^2 x  \nonumber\\ 
&& + (2\epsilon/\omega-1) (E \omega + g^2 + \omega^2/2) \coth x \, {\mathrm{csch}} \,x , 
\label{full2}
\end{eqnarray}

It should be noted that the energy $E$ appearing in these equations is now the regular energy of the AQRM, for the common set of parameter values. 
This establishes the full spectral equivalence between the two systems.
For the QES exceptional values, $E_\pm = n \omega - g^2/\omega \pm \epsilon$, the above results reduce to those given in \S 2.3.

\subsection{Symmetric quantum Rabi model}

We now illustrate this equivalence further for the special case of the symmetric quantum Rabi model when $\epsilon=0$.
The eigenspectrum of the symmetric quantum Rabi model is shown in Figure \ref{rabi0} for a particular set of parameter values. 
The energy levels ${\cal E}$ given by (\ref{Ecom}) for the generalised P\"oschl-Teller potentials (\ref{full1}) and (\ref{full2})  are shown in Figure \ref{pt0} for the same 
set of parameter values. 
The analogous crossing points, at which the QES formalism applies, can be clearly observed.

\begin{figure}[t]
\begin{center}
\includegraphics[width=1.0\columnwidth]{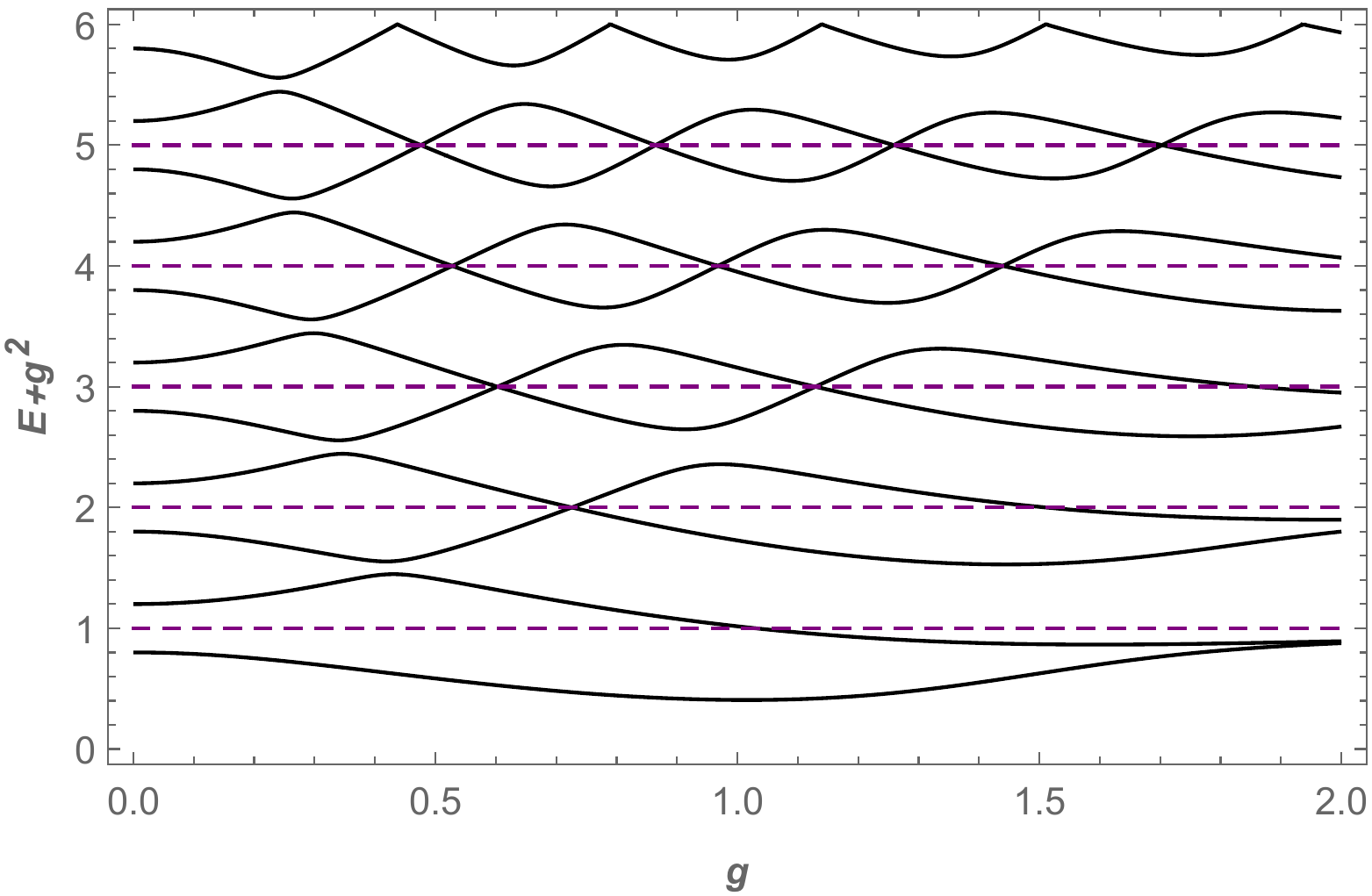}
\caption{Rescaled lowest energy levels $E+g^2$ in the eigenspectrum of the symmetric quantum Rabi model  as a function of the light-matter coupling $g$. 
The parameter values are $\Delta = 1.2$, $\omega = 1$ and $\epsilon = 0$. 
The $n$ crossing points are exactly on the lines $E+g^2=n$ for $n \ge 2$. 
The energy levels have been obtained using Braak's $G$-function  \cite{Braak}.}
\label{rabi0}
\end{center}
\end{figure}

\begin{figure}[h]
\begin{center}
\includegraphics[width=1.0\columnwidth]{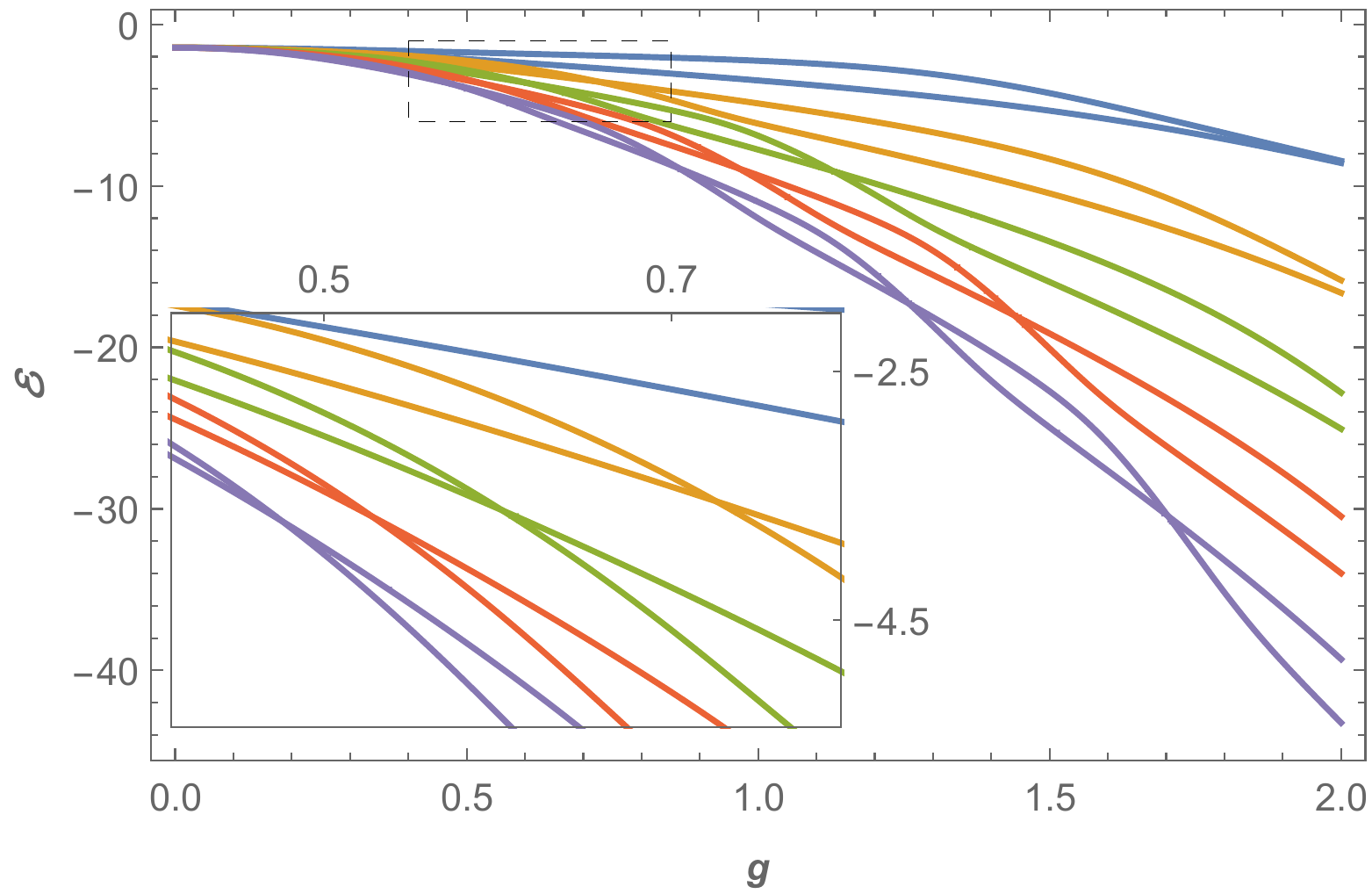}
\caption{The energy levels ${\cal E}$ (\ref{Ecom}) of the generalised P\"oschl-Teller potentials (\ref{full1}) and (\ref{full2}) 
with parameter values $\Delta = 1.2$, $\omega = 1$ and $\epsilon = 0$.
The inset shows a magnification of the indicated region.
The crossing points are the QES exceptional points, with the whole energy spectrum corresponding 
to the energy spectrum of the quantum Rabi model shown in Figure 3.
}
\label{pt0}
\end{center}
\end{figure}

\subsection{Connection to previous results for QES potentials}

We are now in a position to make contact with previous work connecting the quantum Rabi model to a generalised QES P\"oschl-Teller potential \cite{Koc}.
Generalised QES P\"oschl-Teller potentials have also been discussed purely within the QES framework \cite{KK2005}.
In the latter work, the authors begin with the equation
\begin{eqnarray}
 z(1-z) \frac{d^2{\cal R}_j(z)}{d z^2} &+& \left[ L + \case{3}{2} + z(B + 4j- q A^2 z) \right] \frac{d{\cal R}_j(z)}{d z} \nonumber\\
&-&  (\lambda - 2 j q A^2 z) {\cal R}_j(z) = 0 . 
\label{KK}
\end{eqnarray}
Here $L, q, A$ and $\lambda$ are constants with $2j = 0, 1, 2, \ldots$.
First we remark that this equation is precisely equation (\ref{ode1}) subject to the change of variables 
$z=(g+\omega x)/(2 g)$ then multiplying by $-\omega^2$.
We can make the explicit identification 
\begin{eqnarray}
q & =& -{\frac {4{g}^{2}}{{A}^{2}{\omega}^{2}}}, \\
\lambda &=& E^2/\omega^2 - 2 E g^2/\omega^3 + 4g^2 \epsilon /\omega^4 - 3g^4/\omega^4 -\Delta^2/\omega^2 -\epsilon^2/\omega^2 ,\\
B &=& - 1  - 4 g^2/\omega^2 + 2 \epsilon/\omega, \\
2 j & =& E/\omega  + g^2/\omega^2 - \epsilon/\omega, \\
L & =& - E/\omega  - g^2/\omega^2 -1/2 + \epsilon/\omega.
\end{eqnarray}
$A$ is arbitrary, or equivalently $q$ can be taken to be arbitrary and define $A$.
We note that the generalised QES P\"oschl-Teller potentials given in references \cite{Koc} and \cite{KK2005} differ from those derived here, 
because they are based on different transformations compared to (\ref{trans}). 
In this sense our approach follows more closely reference \cite{GKO}, obtaining the same form of generalised QES P\"oschl-Teller potentials derived therein,  
but establishing a spectral equivalence beyond the QES sector with the AQRM.
In the same way the above identification of variables can be used to extend the QES potentials given in \cite{KK2005}, which can now also be related to the AQRM.

\section{Concluding remarks}

Beginning with the Gaudin-like Bethe ansatz equations (\ref{alg1}) and (\ref{alg2}) associated with the QES exceptional points of the AQRM we 
established a spectral equivalence with QES hyperbolic Schr\"odinger potentials on the line, for which similar algebraic 
Bethe ansatz equations were known \cite{DHL}.
This involved generalised QES P\"oschl-Teller potentials of the type (\ref{QESV1}) and (\ref{QESV2}).
Both systems share the same set of constraint polynomials defining the QES exceptional points. 
In this way recent progress on understanding the crossing points in the energy spectrum of the AQRM 
when $\epsilon/\omega \in \case12 \mathbb{Z}$  \cite{Wakayama1,Wakayama2} 
also applies to the energy spectrum of the QES P\"oschl-Teller potentials. 
Here we have been able to write the polynomials $Q_k$ in the form (\ref{Qresult}) in terms of the Gaudin-like Bethe ansatz roots $\{z_k\}$.
The QES spectral equivalence was then extended to the complete spectral equivalence between the AQRM and the 
generalised P\"oschl-Teller potentials (\ref{full1}) and (\ref{full2}). 
The analytic solution of the AQRM thus equally applies to the generalised P\"oschl-Teller potentials.
Given this equivalence between the two systems, it is not unreasonable to expect that the physics of the 
generalised P\"oschl-Teller potentials, and possibly other Schr\"odinger potentials, 
may also be explored in experimental realisations of the quantum Rabi model.\footnote{It is also of interest to see if there is some connection with  
tunneling potentials discussed in terms of the oscillator tunneling dynamics of the quantum Rabi model \cite{EG}. We thank the referee for this remark.}

\ack
CD thanks the Centre for Modern Physics at Chongqing University for hospitality during a visit.
MTB gratefully acknowledges support from Chongqing University and the 1000 Talents Program of China. 
This work is also supported by the Australian Research Council through grant DP170104934.


\section*{References}
\begin{thebibliography}{10}

\bibitem{Rabi}Rabi I I 1936 On the process of space quantization {\sl Phys. Rev.} \textbf{49} 324
\nonum Rabi I I 1937 Space quantization in a gyrating magnetic field {\sl Phys. Rev.} \textbf{51} 652

\bibitem{JC} Jaynes E T and Cummings F W 1963 Comparison of quantum and semiclassical radiation theories 
with application to beam maser Proc. IEEE {\textbf 51}, 89 

\bibitem{intro} Braak D, Chen Q-H, Batchelor M T and Solano E 2016 
Semi-classical and quantum Rabi models: in celebration of 80 years {\sl J. Phys. A} \textbf{49} 300301 

\bibitem{review} Xie Q-T, Zhong H-H, Batchelor M T and Lee C-H 2017 
The quantum Rabi model: solution and dynamics {\sl J. Phys. A} \textbf{50} 113001

\bibitem{Braak} Braak D 2011 Integrability of the Rabi model {\sl Phys. Rev. Lett.} \textbf{107} 100401

\nonum 
Braak D 2013 A generalized $G$-function for the quantum Rabi model {\sl Ann. Phys. (Berlin)} \textbf{525} L23


\bibitem{Heun2} Zhong H, Xie Q, Guan X-W, Batchelor M T, Gao K and Lee C 2014 
Analytical energy spectrum for hybrid mechanical systems {\sl J. Phys. A} \textbf{47} 045301 

\bibitem{hybrid} Treutlein P, Genes C, Hammerer K, Poggio M and Rabl P, in 
{\em Cavity Optomechanics}, Aspelmeyer M, Kippenberg T J and Marquardt F (Eds.) (Springer-Verlag, Berlin, 2014) p 327

\bibitem{cQED} Niemczyk T,  Deppe F,  Huebl H,  Menzel E P,  Hocke F, Schwarz M J,  Garcia-Ripoll J J,  Zueco D,  H\"{u}mmer T, Solano E,
 Marx A and  Gross R 2010 
Circuit quantum electrodynamics in the ultrastrong-coupling regime
 Nat. Phys. \textbf{6} 772


\bibitem{Chen} Chen Q-H, Wang C, He S, Liu T and Wang K-L 2012 
Exact solvability of the quantum Rabi model using Bogoliubov operators {\sl Phys. Rev. A} \textbf{86} 023822

\bibitem{others2} Maciejewski A J, Przybylska M and Stachowiak T 2014 
Analytical method of spectra calculations in the Bargmann representation {\sl Phys. Lett. A}  \textbf{378} 3445 

\bibitem{Judd} Judd B R 1979 Exact solutions to a class of Jahn-Teller systems {\sl J. Phys. C} \textbf{12} 1685

\bibitem{LB1} Li Z-M and Batchelor M T 2015 
Algebraic equations for the exceptional eigenspectrum of the generalized Rabi model 
{\sl J. Phys. A} \textbf{48} 454005 

\nonum Li Z-M and Batchelor M T 2016  Addendum to `Algebraic equations for the exceptional eigenspectrum of the generalized Rabi model'
{\sl J. Phys. A} \textbf{49} 369401

\bibitem{Wakayama1} Wakayama M 2017
Symmetry of asymmetric quantum Rabi models
{\sl J. Phys. A} \textbf{50} 174001

\bibitem{Wakayama2} Kimoto K, Reyes-Bustos C and Wakayama M 2017 
Determinant expressions of constraint polynomials and the spectrum of the asymmetric quantum Rabi model, 
arXiv:1712.04152

\bibitem{BLZ} Batchelor M T, Li Z-M and Zhou H-Q 2016 
Energy landscape and conical intersection points of the driven Rabi model
{\sl J. Phys. A} \textbf{49} 01LT01

\bibitem{Koc} Ko\c{c} R, Koca M and T\"{u}t\"{u}nk\"{u}ler H 2002 Quasi exact solution of the Rabi Hamiltonian {\sl J. Phys. A} \textbf{35} 9425

\bibitem{Zhang} Zhang Y-Z 2013 On the solvability of the quantum Rabi model and its
2-photon and two-mode generalizations {\sl J. Math. Phys.} \textbf{54} 102104 

\bibitem{BZ} Batchelor M T and Zhou H-Q 2015 Integrability versus 
exact solvability in the quantum Rabi and Dicke models {\sl Phys. Rev. A} \textbf 91 053808 

\bibitem{QESa} Turbiner A V 1988
Quasi-exactly solvable problems and $sl(2)$ algebra 
 {\sl Commun. Math. Phys.} \textbf{118} 467

\bibitem{QESb} Ushveridze A G 1993 {\sl Quasi-Exactly Solvable Models in Quantum Mechanics} (Bristol: Institute of Physics Publishing)

\bibitem{DHL} Dunning C, Hibberd K E and Links J 2008 
Some spectral equivalences between Schr\"odinger operators 
{\sl J. Phys. A} \textbf{41}  315211

\bibitem{PT} P\"oschl G and Teller E 1933 
Bemerkungen zur Quantenmechanik des anharmonischen Oszillators 
{\sl Z. Physik} \textbf{83} 143 

\bibitem{HP2017} Hartmann R R and Portnoi M E 2017 
Two-dimensional Dirac particles in a P\"oschl-Teller waveguide 
{\sl Scientific Reports} \textbf{7} 11599 

\bibitem{Schw} Schweber S 1967 On the application of Bargmann Hilbert spaces to dynamical problems {\sl Ann. Phys., NY} \textbf{41} 205

\bibitem{Kus} Ku\'s M 1985 On the spectrum of a two-level system {\sl J. Math. Phys.}  {\bf 26} 2792

\bibitem{GKO} Gonz\'alez-L\'opez A, Kamran N and Olver P J 1993 Normalizability of one-dimensional quasi-exactly solvable Schr\"odinger operators 
{\sl Comm. Math. Phys.} \textbf{153}  117

\bibitem{FGR} Finkel F, Gonz\'alez-L\'opez A and Rodr\'iguez M A 1996 Quasi-exactly solvable potentials on the line and orthogonal polynomials 
{\sl J. Math. Phys.} \textbf{37}  3954

\bibitem{KK2005} Ko\c{c} R and Koca M 2005 
A unified treatment of quasi-exactly solvable potentials I, 
arxiv.org:math-ph/0505002

\bibitem{EG} Irish E K and Gea-Banacloche J 2014 
Oscillator tunneling dynamics in the Rabi model 
{\sl Phys. Rev. B} \textbf{89}  085421 





\end{thebibliography}
\end{document}